\documentclass[12pt]{article}     

\usepackage{graphicx}
\usepackage{amsmath}
\usepackage{mathptmx}  
\usepackage{hyperref}    
\usepackage{enumerate}
\usepackage[misc]{ifsym}
\usepackage{natbib} 
\setcitestyle{authoryear,open={},close={}} 

\usepackage[none]{hyphenat} 

\begin{document}

\thispagestyle{empty}

\noindent {\bf \Large Non-Homogeneity Estimation and Universal Kriging on the Sphere} \newline

\noindent Authors: Nicholas W. Bussberg$^{*,\mbox{ }a}$ (\href{mailto:nbussberg@elon.edu}{nbussberg@elon.edu}) \newline
\indent Jacob Shields$^b$ (\href{mailto:jacob.shields@elancoah.com}{jacob.shields@elancoah.com}) \newline
\indent Chunfeng Huang$^c$ (\href{mailto:huang48@indiana.edu}{huang48@indiana.edu}) \newline

\noindent $^{*}$ Corresponding author.\newline

\noindent $^a$ Elon University, Department of Mathematics and Statistics, 100 Campus Drive, Elon, NC 27244 \newline
\noindent $^b$ Elanco Animal Health, 2500 Innovation Way, Greenfield, IN 46140 \newline
\noindent $^c$ Indiana University, Department of Statistics, Informatics East, 919 E 10th St, Bloomington, IN 47408 \newline

\begin{abstract}
Kriging is a widely recognized method for making spatial predictions. 
On the sphere, popular methods such as ordinary kriging assume that the spatial process is intrinsically homogeneous. 
However, intrinsic homogeneity is too strict in many cases. 
This research uses intrinsic random function (IRF) theory to relax the homogeneity assumption. 
A key component of modeling IRF processes is estimating the degree of non-homogeneity. 
A graphical approach is proposed to accomplish this estimation. With the ability to estimate non-homogeneity, an IRF universal kriging procedure can be developed. 
Results from simulation studies are provided to demonstrate the advantage of using IRF universal kriging as opposed to ordinary kriging when the underlying process is not intrinsically homogeneous. \newline

\end{abstract}

\newpage


\section{Introduction}
\label{sec:intro}

Since the late 1980's, the field of spatial statistics has expanded into almost every area of applied research. Researchers use spatial models to analyze global temperatures, determine pubic health and safety, and investigate patterns in the brain. Much research has been done in Euclidean spaces (e.g., \cite{Cressie1993, Stein1999, ChilesDelfiner2012}), but processes on the spherical domain have gained more attention recently (e.g., \cite{HuangZhang2011, Gneiting2013, PorcuBevilacqua2016}). 

Choosing spherical models is naturally important when modeling processes on the earth. Using models developed in Euclidean space is appropriate on small scales where the curvature of the earth does not largely impact the process (\cite{JeongJun2015}). However, if these models are used on the whole sphere, models should be designed with the spherical geometry in mind, and great circle distances should be used instead of Euclidean distances (\cite{HuangZhang2011}). 

In addition to using spherical models, many models also assume homogeneous or intrinsically homogeneous processes, which is often deemed unrealistic in practice. Several approaches have been proposed to relax this assumption such as axial symmetry (\cite{Jones1963, Stein2007, HuangZhang2012}) and kernel convolution (\cite{ZhuWu2010, HeatonKatzfuss2014}). 
\citeauthor{HuangZhang2019} (\citeyear{HuangZhang2019}) extend the notion of intrinsic random functions (IRF) (\cite{Matheron1973}) to the spherical domain, where monomials in Euclidean spaces annihilated by the allowable measure are replaced by spherical harmonics on the sphere. The theoretical properties of IRFs on the sphere are developed, and the connection with the reproducing kernel Hilbert space is discussed. In particular, \citeauthor{HuangZhang2019} (\citeyear{HuangZhang2019}) show that an IRF on the sphere is characterized by its frequency-truncated process.

Based on this theoretical development, this paper provides a method for estimating the degree of non-homogeneity. Theorem 1 in \citeauthor{HuangZhang2019} (\citeyear{HuangZhang2019}) shows that a random process on the sphere is IRF$\kappa$ if and only if its low-frequency truncated process is homogeneous. Based on this findings, a graphical tool for estimating $\kappa$, which is the degree of non-homogeneity on the sphere, is developed in Sect.~\ref{sec:kest}. 

Once the degree of non-homogeneity is estimated, it is fairly straightforward to carry out universal kriging. The procedure for universal kriging is presented in Sect.~\ref{sec:uk}. Note that \citeauthor{Matheron1973} (\citeyear{Matheron1973}) discussed universal kriging based on the intrinsic random functions, where various order of monomials can be modeled while their coefficients do not need to be estimated in the kriging formula. For example, ordinary kriging is to assume intrinsic stationarity with a constant mean function, and such constant does not play a role in variogram-based kriging formula. Higher order kriging beyond ordinary kriging is less common in practice. 
To the authors best knowledge, a method for estimating the degree of non-stationarity in Euclidean spaces has yet to be developed. However, on the sphere, the degree of non-homogeneity of an IRF can be estimated. This paves a way for a truly universal kriging as Matheron envisioned in 1973. 

Then, in Sect.~\ref{sec:sim}, results from simulation studies are presented to show IRF-based universal kriging can be applied, including the estimate of the degree of non-homogeneity. Universal kriging is presented in comparison to ordinary kriging and is shown to be advantageous when the underlying data are not intrinsically homogeneous.


\section{Degree of Non-Homogeneity Estimation}
\label{sec:kest}

\citeauthor{HuangZhang2019} (\citeyear{HuangZhang2019}) introduced the notion of IRF on the sphere and revealed the unique theoretical properties of the IRF on the sphere. In particular, a real-valued continuous (in quadratic mean) random function $Z(x), x \in S^2$ is an IRF if and only if its low frequency-truncated process $Z_{\kappa}(x)$ is homogenous. Here, 
\[
Z(x) = \sum_{l=0}^{\infty} \sum_{m=-l}^l Z_{l,m} Y_l^m(x),
\]
and
\[
Z_{\kappa}(x) = \sum_{l=\kappa}^{\infty} \sum_{m=-l}^l Z_{l,m} Y_l^m(x),
\]
where the real-valued spherical harmonic functions are 
\[
\left\{ \begin{array}{l} Y_l^m(x) = \sqrt{ \frac{ 2l+1}{2 \pi} \frac{ (l-m)!}{(l+m)!}} P_l^m(\cos \zeta) \cos (m \psi), \quad m=1, \ldots, l, \\
Y_l^0(x) = \sqrt{ \frac{2l+1}{4\pi} } P_l^0(\cos \zeta), \\
Y_l^{-m}(x) = \sqrt{ \frac{2l+1}{2 \pi} \frac{ (l-m)!}{(l+m)!}} P_l^m(\cos \zeta) \sin (m \psi), \quad m=1, \ldots, l, \end{array} \right.
\]
$P_l^m(\cdot)$ are the associated Legendre polynomials, $P_l^0(\cdot) \equiv P_l(\cdot)$ are the Legendre polynomials, and $x = (\psi, \zeta)$ with longitude $\psi \in [0, 2\pi)$ and the latitude $\zeta \in [0, \pi]$. The coefficients $Z_{l,m}$ are random variables
\[
Z_{l,m} = \int_{S^2} Z(x) Y_l^m(x) dx.
\]

Then, for this homogenous process $Z_{\kappa}(\cdot)$, the coefficients are uncorrelated (\cite{Obukhov1947})
\[
\mbox{cov} (Z_{l,m}, Z_{l',m'}) = a_l I(l,l') I(m,m'),
\]
where $l, l' \ge \kappa, |m|, |m'| \le l$ and $a_l \ge 0$. Using this form, the associated covariance function of the process $Z_{\kappa}(\cdot)$ is then
\[
\phi_{\kappa}(d(x,y)) = \mbox{cov} (Z_{\kappa}(x), Z_{\kappa}(y)) = \sum_{l=\kappa}^{\infty} \frac{2l+1}{4 \pi} a_l P_l (\cos (d(x,y))),
\]
where $d(x,y)$ is the spherical distance between $x, y \in S^2$. 
This function plays the important role of the generalized covariance for IRF theory (\cite{Matheron1973}), and is termed as the intrinsic covariance function (ICF) in this manuscript. 

To model processes as IRF$\kappa$, both $\kappa$ and $\phi_{\kappa}$ must be estimated. 
Based on Theorem 1 in \citeauthor{HuangZhang2019} (\citeyear{HuangZhang2019}), the low frequency-truncated process $Z_{\kappa}(x)$ is homogeneous. Therefore, the further truncated process $Z_{\kappa+1}(x)$ is also homogenous with 
\[
\phi_{\kappa+1}(d(x,y)) = \sum_{l=\kappa+1}^{\infty} \frac{2l+1}{4 \pi} a_l P_l (\cos (d(x,y))).
\]
The difference of these two ICFs is
\[
\phi_{\kappa}(d(x,y)) - \phi_{\kappa+1}(d(x,y)) = \frac{2 \kappa+1}{4 \pi} a_{\kappa} P_{\kappa} (\cos (d(x,y))).
\]
This remains true for any additional truncation. In general, 
\[
\phi_j (d(x,y)) - \phi_{j+1} (d(x,y)) = \frac{2 j+1}{4 \pi} a_j P_j(\cos(d(x,y))), \quad j=\kappa, \kappa+1, \ldots.
\]
Because $a_j$ is a constant, then
\[
\phi_j(d(x,y)) - \phi_{j+1} (d(x,y)) \propto P_j (\cos (d(x,y))), \quad j \ge \kappa.
\]

Based on this discussion, for any spatial lag $h_0 = 0 < h_1 < \ldots < h_m \le \pi$, 
\[
\left( \phi_j(h_i) - \phi_{j+1} (h_i) - \frac{2 j+1}{4 \pi} a_j P_j (\cos h_i) \right) = 0, \quad j=\kappa, \kappa+1, \ldots, \quad i=0, 1, \ldots, m.
\]
Therefore, if the estimates $\Big\{\hat{\phi}_j(h_i)\Big\}_{j=\kappa, \kappa+1, \ldots}^{i=0, 1, \ldots, m}$ could be obtained, one would expect that 
\begin{align}
\sum_{i=1}^m \left( \hat{\phi}_j(h_i) - \hat{\phi}_{j+1} (h_i) - \frac{2j+1}{4\pi} a_j P_j (\cos h_i) \right)^2 \label{eq:small-phi}
\end{align}
to be small and remain small for all $j \ge \kappa$. 

To find $\hat{\phi}_j(\cdot)$, a method of moment estimator can be constructed. The full spatial process $Z(x)$ can be expanded as
\[
Z(x) = \sum_{l < j} \sum_{m=-l}^l Z_{l,m}Y_l^m(x) + Z_j(x),
\]
where $Z_j(x)$ is the $j$-truncated process. When an IRF$\kappa$ $Z(x)$ is observed, the low-frequency truncated process can be approximated by the residual process $Z_{j,r}(x)$ of $Z(x)$ regressed on $\{ Y_l^m(\cdot) \}_{l < j}$. When $j \ge \kappa$, $Z_j(x)$ is homogenous with ICF $\phi_j(\cdot)$. Therefore, a direct method of moment estimator is
\[
\hat{\phi}_j(h_i) = \frac{1}{ | N_{h_i}|} \sum_{(x,y) \in N_{h_i}} Z_{j,r}(x) Z_{j,r}(y),
\]
where $N_{h_i}$ is the set of all pairs $(x,y)$ such that $d(x,y) = h_i$ and $|N_{h_i}|$ is the cardinality of this set. 
For $h_0, h_1, \ldots, h_m$, this estimate for $\hat{\phi}_j(\cdot)$ satisfies Eq.~(\ref{eq:small-phi}) such that Eq.~(\ref{eq:small-phi}) shall be small and remain small for all $j \ge \kappa$. 

Note that the summation in Eq.~(\ref{eq:small-phi}) requires an estimate for $a_j$. This is readily available in $\hat{\phi}_j(\cdot)$ by solving 
\[
\phi_j(0) - \phi_{j+1}(0) = \frac{2j+1}{4 \pi} a_j , \quad j \ge \kappa.
\]
This is because the Legendre polynomial $P_j(\cos(0)) = 1$ for $d(x,y) = 0$. An estimate for $a_j$ could then be derived as 
\[
\hat{a}_j = \frac{4\pi}{2j+1}(\hat{\phi}_j(0) - \hat{\phi}_{j+1}(0)).
\]

Combining this with Eq.~(\ref{eq:small-phi}) yields the form
\begin{align}
\sum_{i=1}^m \left( \hat{\phi}_j(h_i) - \hat{\phi}_{j+1}(h_i) - \{ \hat{\phi}_j(0) - \hat{\phi}_{j+1}(0) \} P_j (\cos (h_i)) \right)^2 
\label{eq:small-phi0}
\end{align}

Using Eq.~(\ref{eq:small-phi0}), a criterion for estimating $\kappa$ can be established. It is first important to recognize that, for $j < \kappa$ (i.e., prior to the removal of the spherical harmonics of order less than $\kappa$), the residual processes are non-homogeneous. In fact, they are governed by a reproducing kernel, which includes the ICF and other terms based on choices of arbitrary chosen points on the sphere (\cite{HuangZhang2019}). 
Thus, when $j < \kappa$, the notation $\hat{\phi}(\cdot)$ is not appropriate. A new notation $G(\cdot)$ can be defined as 
\begin{align}
G(j, h_i) = \frac{1}{|N_{h_i}|} \sum_{(x,y) \in N_{h_i}} Z_{j,r}(x) Z_{j,r}(h).
\label{eq:Gj}
\end{align}
When $j < \kappa$, $G(j, h_i)$ is meaningless, and
\begin{align}
\mathcal{M}(j) = \sum_{i=1}^m \left( G(j, h_i) - G(j+1, h_i) - \{ G(j,0) - G(j+1, 0) \} P_j(\cos (h_i)) \right)^2
\label{eq:Mj}
\end{align}
can be completely arbitrary. However, when $j \ge \kappa$,
\[
G(j, h_i) = \hat{\phi}_j(h_i)
\]
(i.e., $G(j, h_i)$ is an estimator for the ICF), and then $\mathcal{M}(j)$ shall be small and remain small. 

\begin{figure}
	\includegraphics[width=0.75\textwidth]{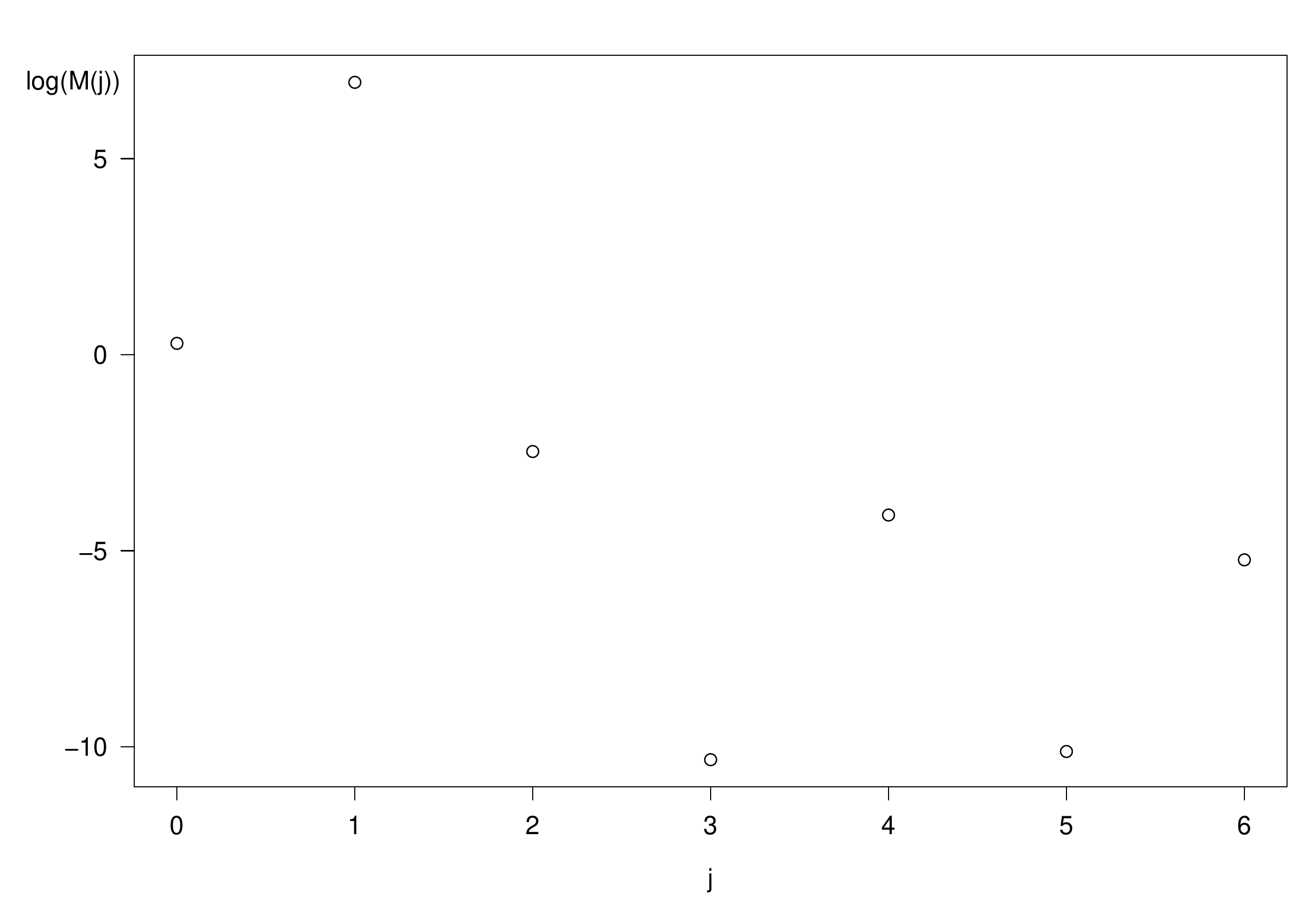}
	\caption{Criterion plot of $\log(\mathcal{M}(j))$ vs. varying values of $j$ for a simulated IRF2 process.}
	\label{fig:critplot2}
\end{figure}

\begin{figure}
	\includegraphics[width=0.75\textwidth]{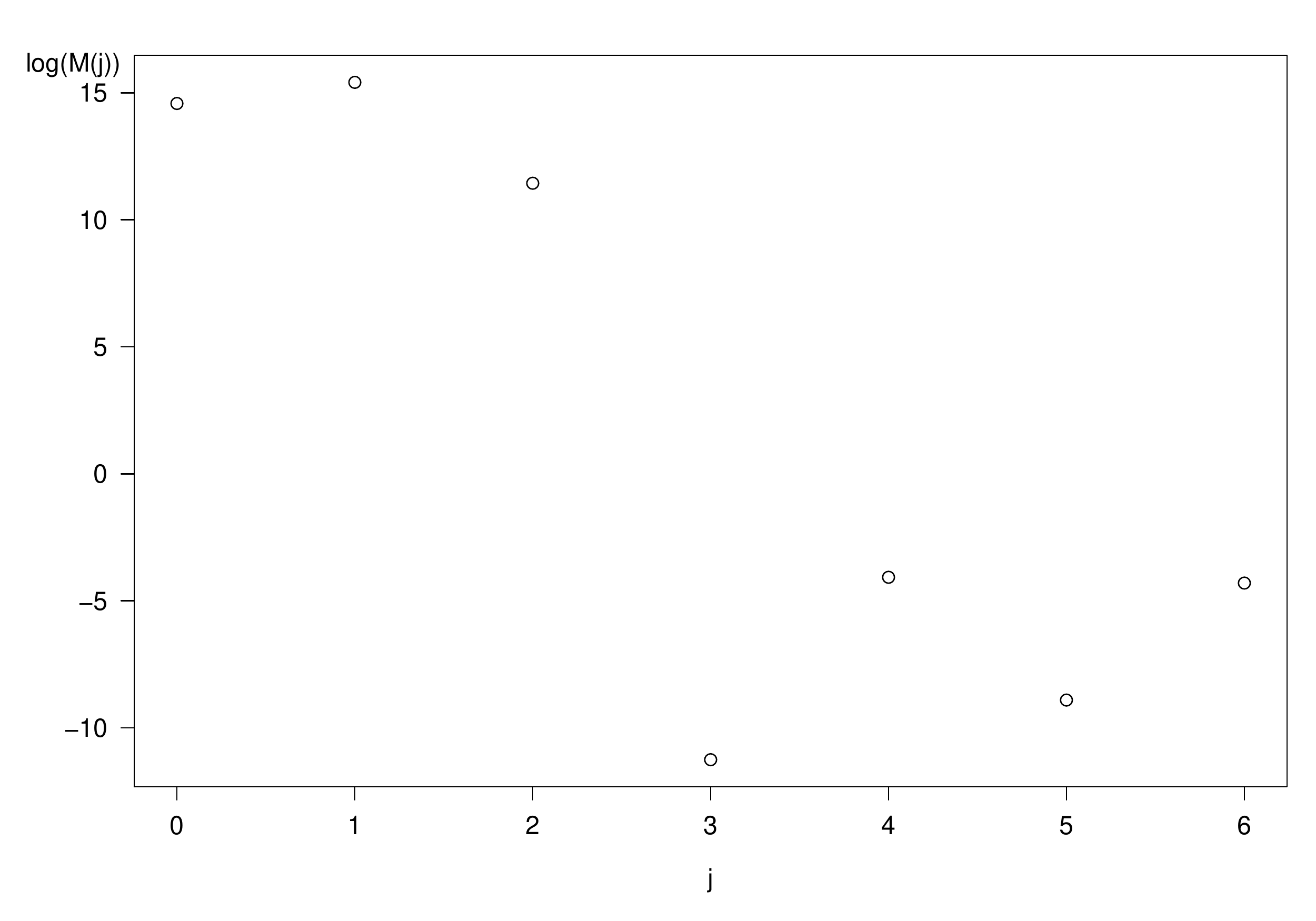}
	\caption{Criterion plot of $\log(\mathcal{M}(j))$ vs. varying values of $j$ for a simulated IRF3 process.}
	\label{fig:critplot3}
\end{figure}

The criterion $\mathcal{M}(\cdot)$ can thus be used to construct a graphical procedure for estimating the degree of non-homogeneity, $\kappa$. 
Based on the previous discussion of $\mathcal{M}(j)$ in Eq.~(\ref{eq:Mj}), it is clear that plotting $\mathcal{M}(j)$ against $j$ should provide an estimate for $\kappa$. 
The estimate for $\kappa$ will be the order $j$ when $\mathcal{M}(j)$ becomes and remains small. However, it can take arbitrary values before $j$ reaches the right order. For example, Figures \ref{fig:critplot2} and \ref{fig:critplot3} shows the results from a simulated IRF$\kappa$, where $\kappa=2$ and 3. (Details of the simulation setup is in Sect.~\ref{sec:sim}. Note: the log of $\mathcal{M}(j)$ was taken in the graphs.)
In the IRF2 plot, $\mathcal{M}(j)$ is small for all $j \ge 2$, and is bigger in magnitude for $j=0,1$. 
Thus, the estimate for $\kappa$ would clearly be $\hat{\kappa}=2$. The same pattern appears in the second plot, where an IRF$3$ is observed.

\section{Universal Kriging}
\label{sec:uk}

Kriging is the most commonly used prediction method in spatial statistics. \citeauthor{Matheron1973} (\citeyear{Matheron1973}) developed the universal kriging through the IRF theory in Euclidean spaces. This approach is extended to the sphere in \citeauthor{HuangZhang2019} (\citeyear{HuangZhang2019}). With the degree of non-homogeneity estimated in Sect.~\ref{sec:kest}, a practical universal kriging procedure can be developed. 
Following \citeauthor{HuangZhang2019} (\citeyear{HuangZhang2019}), first assume that data $\{ (x_i, w_i), i=1, \ldots, n\}$ is observed from an IRF$\kappa$ process $Z(x)$, where
\[
w_i = Z(x_i) + \epsilon_i,
\]
$\epsilon_i$ are uncorrelated measure errors with mean $0$ and variance $\sigma^2$, $Z(x)$ is an IRF$\kappa$ process with an ICF $\phi_{\kappa}(\cdot)$ and a mean function with unknown coefficients $\beta_{l,m}$ being $\mbox{E} [Z(x)] = \sum_{l < \kappa} \sum_{m=-l}^l \beta_{l,m}  Y_l^m (x)$. 
Then, universal kriging at an unsampled location $x_0$ is $\hat{Z}(x_0) = \eta^T w$, where $w = (w_1, \ldots, w_n)^T$ and the coefficients $\eta=(\eta_1, \ldots, \eta_n)^T$ can be solved through
\[
\left\{ \begin{array}{l} (\Psi + \sigma^2 I) \eta + Q \rho = \phi, \\ 
Q^T \eta = q, \end{array} \right.
\label{eq:dualkrg}
\]
where $\Psi = \{ \phi_{\kappa} (d(x_i, x_j)) \}_{i,j=1,\ldots,n}$, $\phi=(\phi_{\kappa}(d(x_1,x_0)), \ldots, \phi_{\kappa}(d(x_n,x_0)))^T$, $\rho$ is the Lagrange multiplier of length $\kappa^2$, $Q=\{q_{\nu}(x_i)\}_{i=1,\ldots,n,\nu=1,\ldots,\kappa^2} \}$, $q=(q_1(x_0), \ldots, q_{\kappa^2}(x_0))^T$. Here, $\{ q_{\nu}(x), \nu=1,\ldots, \kappa^2\}$ is a newly arranged lower spherical harmonics $\{ Y_l^m(x), l < \kappa, |m| \le l\}$ for simplicity. 

Combining this derivation with $\kappa$ estimation presented in the previous section, a practical universal kriging procedure can be formed (Algorithm 1). In this algorithm, a positive definite ICF needs to be estimated. To complete the universal kriging algorithm, the weighted least square procedure for parametric modeling is adopted (\cite{Cressie1985}). 
Let $\phi_{\kappa}(h, \theta)$ be a parametric ICF of order $\kappa$, where $h$ is the lag and $\theta$ are the parameters. While generalized least squares or ordinary least squares can be applied, \citeauthor{Cressie1985} (\citeyear{Cressie1985}) proposed the weighted least squares approach as a good compromise between statistical efficiency and computational demands. That is, to find the estimate, minimize
\[
\sum_{i=0}^m |N_{h_i}| \left( \frac{G(\kappa, h_i)}{\phi_{\kappa} (h_i, \theta)} -1 \right)^2,
\]
where $G(\cdot, \cdot)$ and $|N_{h_i}|$ can be found in Eq.~\ref{eq:Gj}. In Algorithm 1, $\kappa$ is naturally replaced by $\hat{\kappa}$.

\begin{table*}
\label{tab:algorithm}
\begin{tabular}{c l}
	\hline\noalign{\smallskip}
	& Algorithm 1. Universal kriging procedure \\
	\noalign{\smallskip}\hline\noalign{\smallskip}
    	1. & Estimate the order of non-homogeneity, $\hat{\kappa}$. \\
	2. & Estimate the ICF, $\hat{\phi_{\kappa}}(\cdot)$. \\
	3. & Compute $q, Q, \Psi, \phi$ in Eq.~(\ref{eq:dualkrg}). \\
	4. & Obtain $\eta$ to find $\hat{Z}(x_0) = \eta^T w$. \\
	\noalign{\smallskip}\hline
\end{tabular}
\end{table*}


\section{Simulation Study Comparing Ordinary and Universal Kriging on the Sphere}
\label{sec:sim}

In the family of universal kriging, ordinary kriging may be the most commonly used procedure in practice. Ordinary kriging on the sphere assumes intrinsic homogeneity (\cite{HuangZhang2011}), which is IRF with $\kappa=1$ (i.e., IRF1).
Thus, it is prudent to investigate whether IRF universal kriging offers better predictions at unobserved locations when the underlying process is not intrinsically homogeneous.

To simulate a Gaussian IRF$\kappa$ on the sphere, we need its covariance function. The covariance function can be derived utilizing the connection between IRF and RKHS. First, denote nil space for the semi-inner product defined in \citeauthor{HuangZhang2019} (\citeyear{HuangZhang2019}) as $N$, where $\dim(N) = \kappa^2$. Then, the associated reproducing kernel for the covariance function can be written as (\cite{HuangZhang2019}). 
\begin{align}
& H_{\kappa}(x,y) = \phi_{\kappa} (d(x,y)) - \sum_{\nu=1}^{\kappa^2} \{ \phi_{\kappa} (d(x, \tau_{\nu})) p_{\nu}(y) + \phi_{\kappa} (d(y, \tau_{\nu})) p_{\nu}(x) \}  \nonumber \\
& \quad + \sum_{\nu=1}^{\kappa^2} \sum_{\mu=1}^{\kappa^2} \phi_{\kappa} (d(\tau_{\nu}, \tau_{\mu})) p_{\nu}(x) p_{\nu}(y) + \sum_{\nu=1}^{\kappa^2} p_{\nu}(x) p_{\nu}(y),
\label{eq:repker}
\end{align}
where $\{\tau_1, \ldots, \tau_{\kappa^2}\} \in S^2$ and $p_1(\cdot), \ldots, p_{\kappa^2}(\cdot) \in N$ such that $p_{\nu}(\tau_{\mu}) = I(\nu, \mu)$ for $1 \le \nu, \mu \le \kappa^2$. This is clearly a non-homogeneous covariance function. \\

For the simulation studies in this research, the ICFs are simple parametric functions, where we set $a_l = r^l, 0 \le r < 1,$ so that
\[
\phi_{\kappa}(h) = \sum_{l=\kappa}^{\infty} \frac{2l+1}{4 \pi} r^l P_l(\cos h).
\]
This is the truncated version of a well-known model
\[
\sum_{l=0}^{\infty} \frac{2l+1}{4 \pi} r^l P_l(\cos ) = \frac{1-r^2}{4 \pi} (1-2r \cos(h) + r^2)^{-3/2}.
\]
Other parametric models could have be used. In this study, we use this ``simple" parametric function to demonstrate our $\kappa$ estimation and universal kriging methods.

\begin{figure}
	\includegraphics[width=0.75\textwidth]{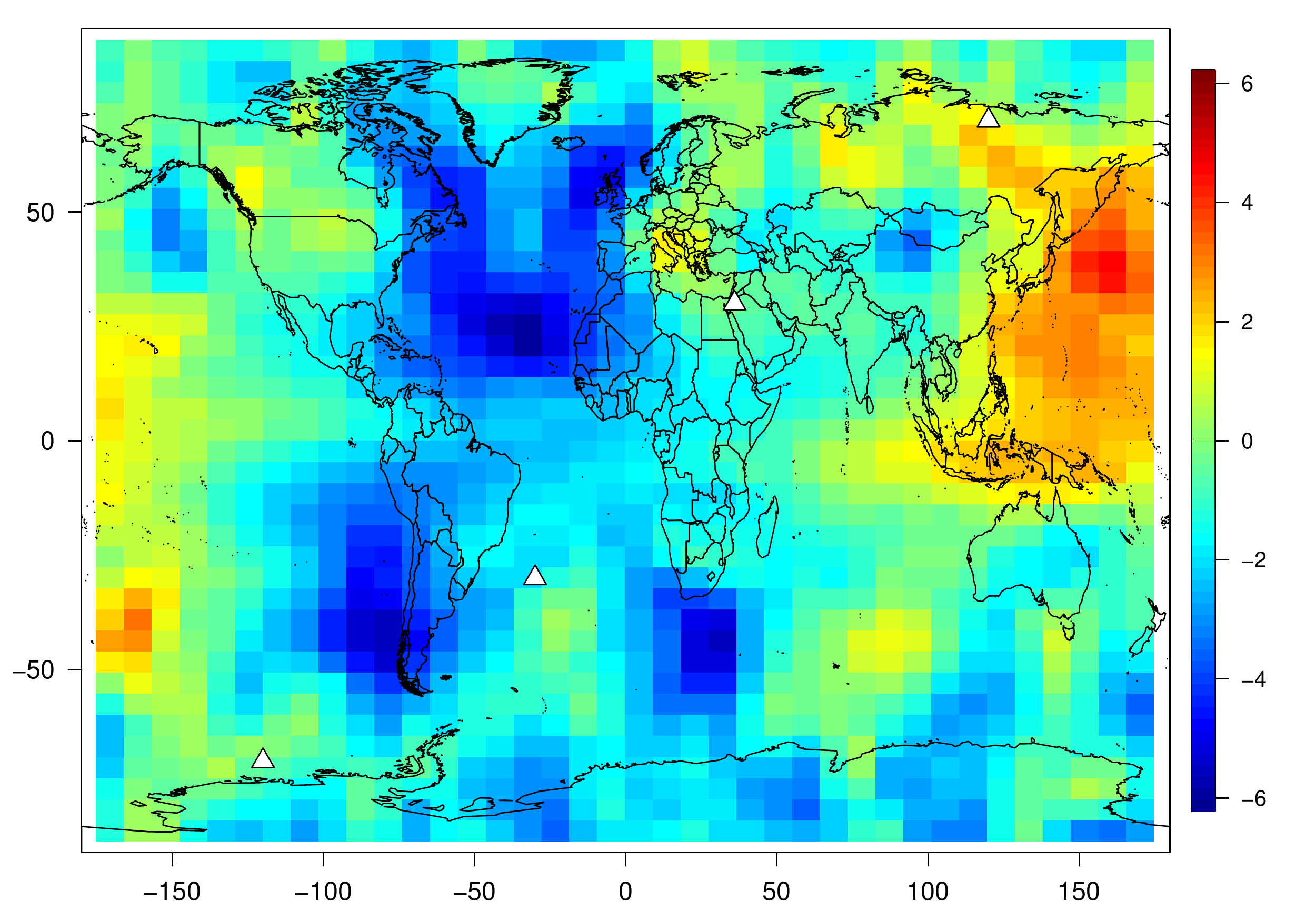}
	\caption{Heat map of an IRF2 simulated processes. The arbitrary locations $\tau$ necessary for the simulation are plotted as white triangles.}
	\label{fig:fullsim2}
\end{figure}

\begin{figure}
	\includegraphics[width=0.75\textwidth]{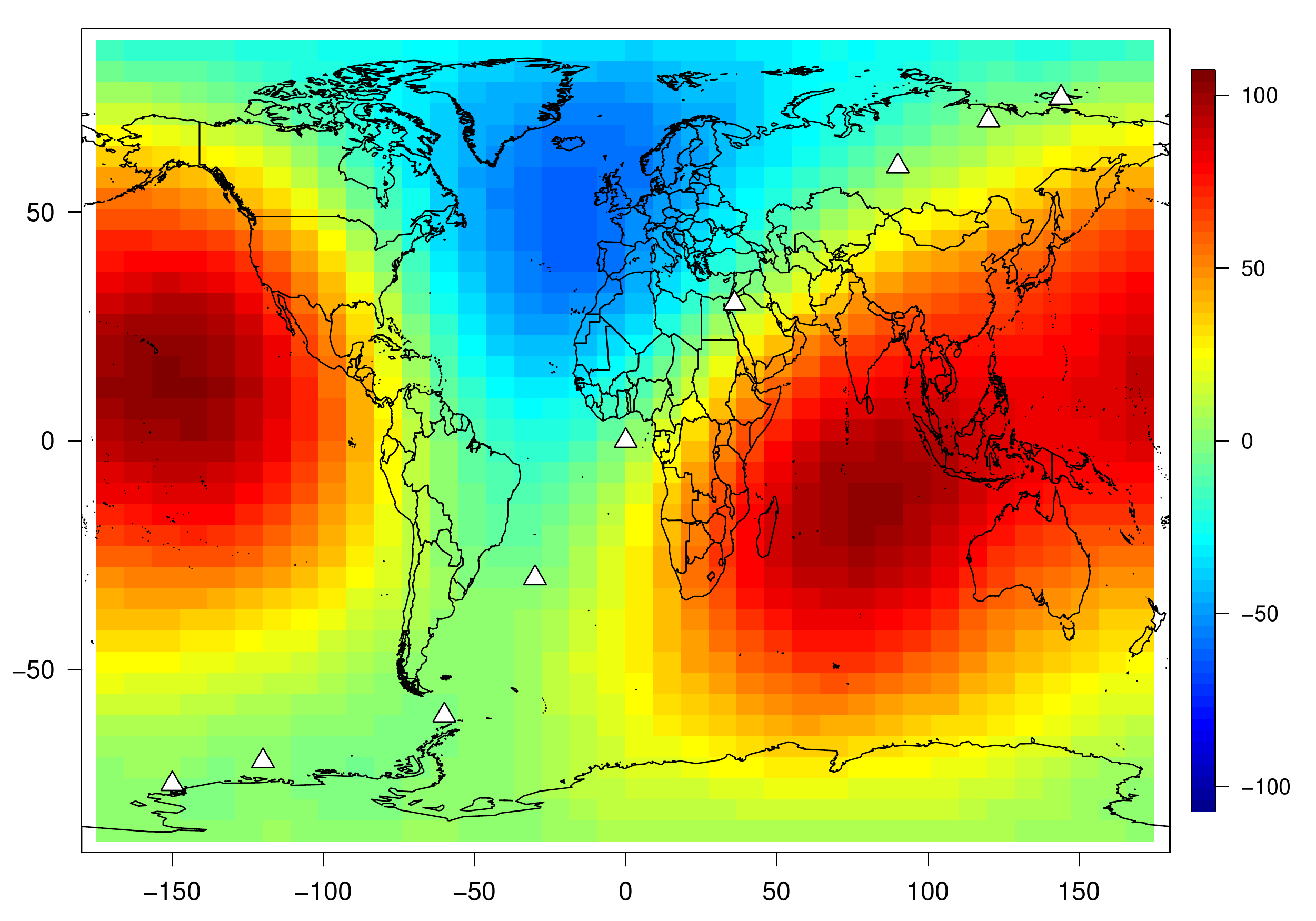}
	\caption{Heat map of an IRF2 simulated processes. The arbitrary locations $\tau$ necessary for the simulation are plotted as white triangles.}
	\label{fig:fullsim3}
\end{figure}

Using this parametric model, IRF processes with $\kappa=2$ and 3 were simulated on 1,500 randomly selected locations on the sphere. The ICFs were both generated with a parameter $r=0.75$. To simulate the full covariance structure, the arbitrary locations $\tau$ from Eq.~(\ref{eq:repker}) need to chosen. For the IRF2 simulations, the $\kappa^2 = 4$ locations were $(\pi/9, \pi/3), (\pi/3, 5\pi/6)$, $(2\pi/3, 6\pi/5)$, and $(8\pi/9, 5\pi/3)$. For the IRF3 simulations, the $\kappa^2 = 9$ locations were $(\pi/12, \pi/6)$, $(\pi/9, \pi/3), (\pi/6, 2\pi/3)$, $(\pi/3, 5\pi/6), (\pi/2, \pi)$, $(2\pi/3, 6\pi/5), (5\pi/6, 3\pi/2)$, $(8\pi/9, 5\pi/3)$, and $(11\pi/12, 9\pi/5)$. 
Figures \ref{fig:fullsim2} and \ref{fig:fullsim3} display the simulated data for IRF2 and IRF3, respectively, with the locations $\tau$ shown as white triangles.

The simulated datasets were split into training and testing subsets so that ordinary kriging and universal kriging predictions could be compared against known values. For both the IRF2 and IRF3 simulations, 90\% of the data were randomly selected 90\% of the data to be the training dataset, and the other 10\% was used as the testing dataset. For the training data, both $\kappa$ and the ICF were estimated. Those values were then used to make predictions at the ``unobserved'' testing locations with kriging.

\begin{figure}
	\includegraphics[width=0.75\textwidth]{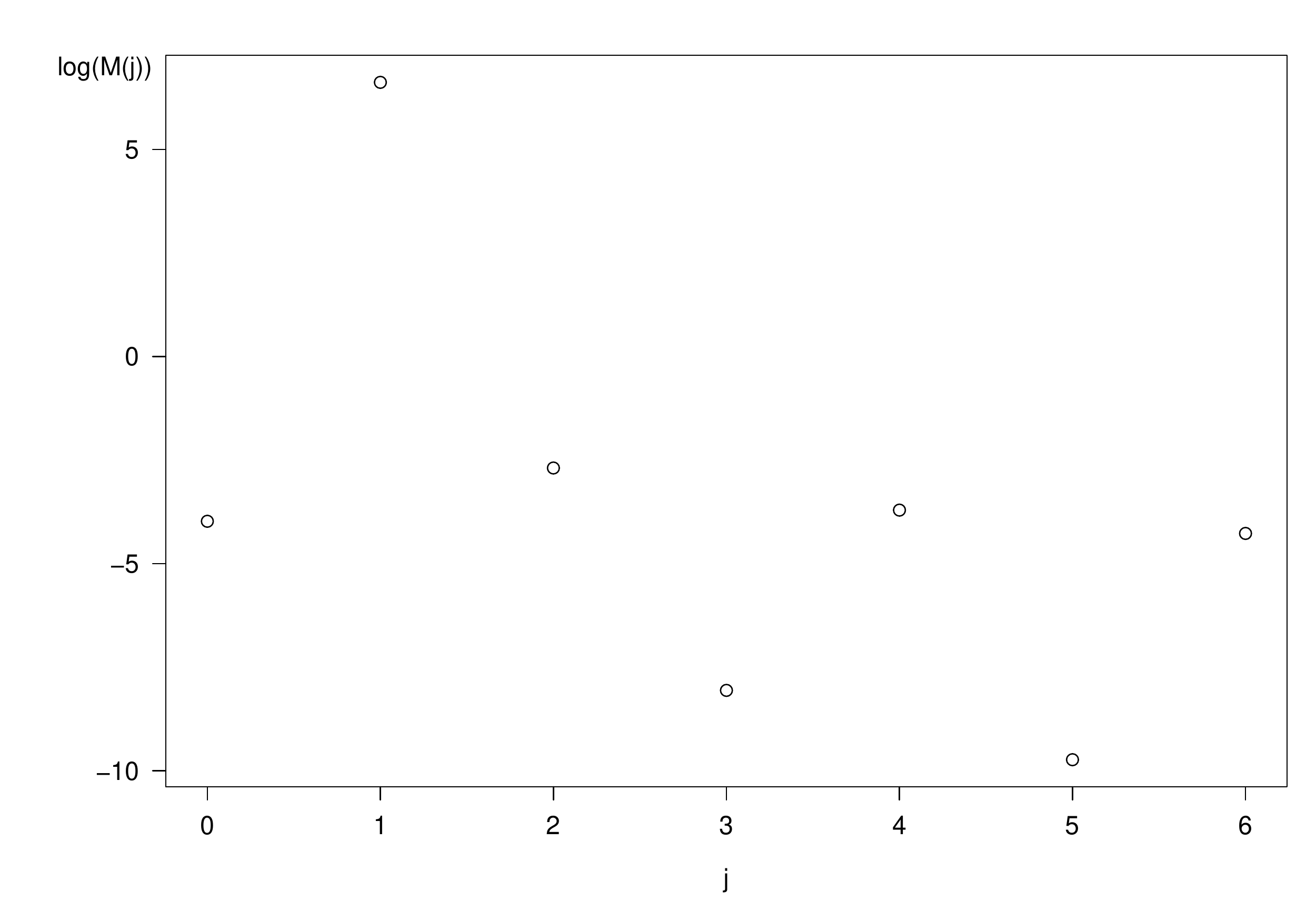}
	\caption{Criterion plot of $\log(\mathcal{M}(j))$ for the simulated IRF2 process used in the kriging simulation study. The plot was generated with the training dataset, which consisted of a randomly selected 90\% of the full dataset.}
	\label{fig:critplots-study2}
\end{figure}

\begin{figure}
	\includegraphics[width=0.75\textwidth]{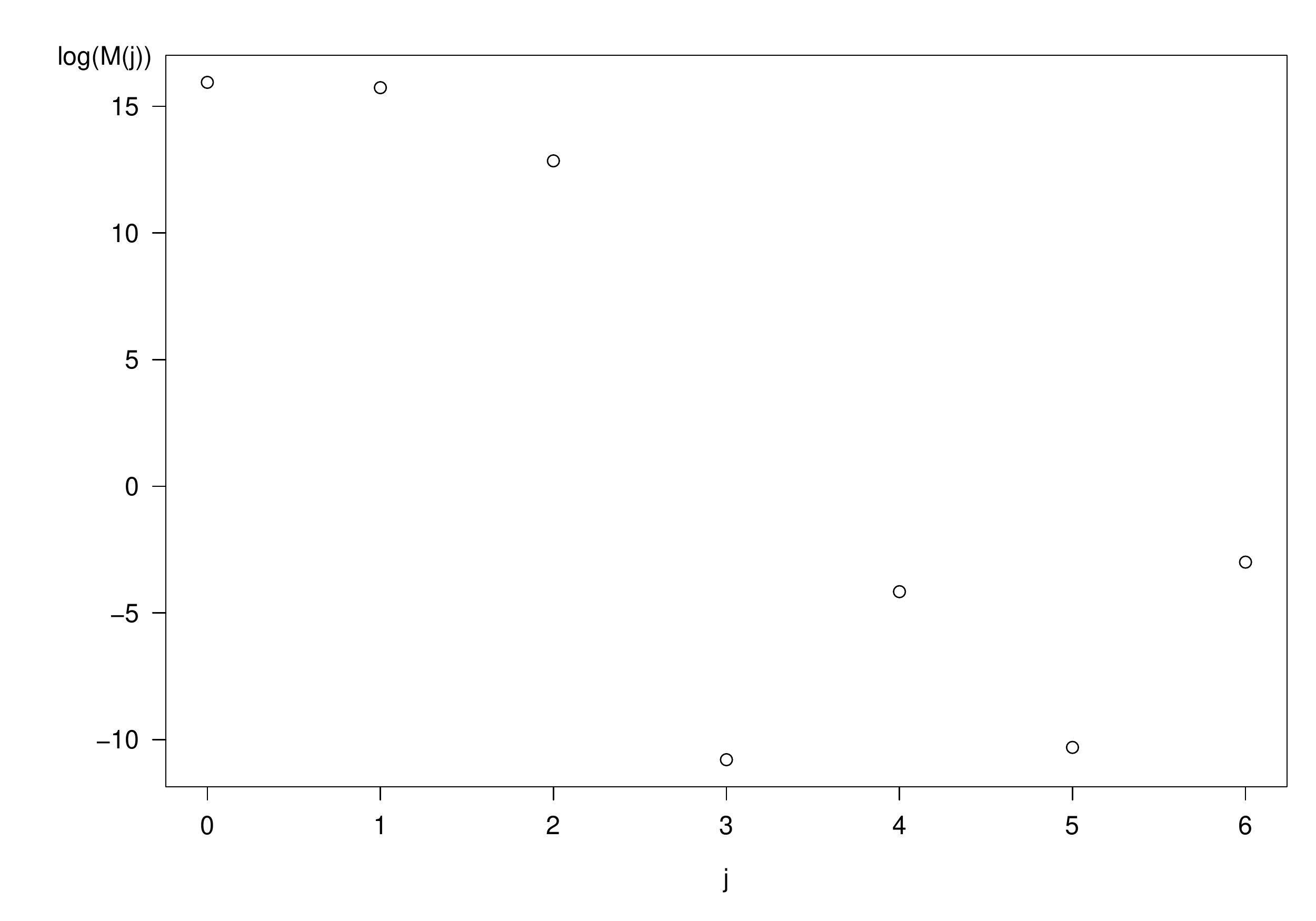}
	\caption{Criterion plot of $\log(\mathcal{M}(j))$ for the simulated IRF3 process used in the kriging simulation study. The plot was generated with the training dataset, which consisted of a randomly selected 90\% of the full dataset.}
	\label{fig:critplots-study3}
\end{figure}

For IRF universal kriging, the degree of non-homogeneity $\kappa$ must first be estimated. 
Using the graphical procedure for both the IRF2 and IRF3 training data subsets, the estimates of $\hat{\kappa}=2$ and 3 were chosen (Figs.~\ref{fig:critplots-study2} and \ref{fig:critplots-study3}, respectively. Note that Fig.~\ref{fig:critplot2} and Fig.~\ref{fig:critplots-study2} are very similar, but Fig.~\ref{fig:critplots-study2} was generated with 90\% of the data whereas Fig.~\ref{fig:critplot2} was created with all of the data. A similar comparison is true for Figs.~\ref{fig:critplot3} and \ref{fig:critplots-study3}). 
For example, in the IRF3 plot, the $\mathcal{M}(j)$ values are large for $j=0, 1,$ and 2, but the values decrease and become relatively stable at $j \ge 3$. This is clearly an indication that the process is neither IRF0, IRF1, or IRF2, and is likely IRF3. Note that in this simulation, the maximum degree $j$ was set at 7. This number can be extended to an arbitrarily larger number. As a maximum value must be chosen for computational purposes, the implied assumption is that the underlying process is an IRF, but $\kappa$ does not exceed the chosen value.

For ordinary kriging, the degree of non-homogeneity is not estimated. The assumption is an intrinsically homogeneous process, which implies that $\hat{\kappa} = 1$. Thus, for this study, $\hat{\kappa}$ was set to 1 for ordinary kriging.

For both ordinary kriging and universal kriging, the weighted least squares approach was used to estimate the parameter $r$, which in turn allows for the estimation of the ICF. For these simulations, $\sigma^2$ was assumed to be 0 to focus on the difference between ordinary kriging and universal kriging in simpler setting. Perhaps unsurprisingly, the estimates of $r$ were closer to the true value of 0.75 when the degree of non-homogeneity was estimated instead of assumed to be 1 (Table \ref{tab:krg-rhat}). The IRF approach had estimates of $\hat{r} = 0.717$ and $0.707$ for IRF2 and IRF3, respectively, where the ordinary kriging estimates were $\hat{r}=0.888$ and $0.992$.

\begin{table}
\caption{Estimates for $\kappa$ and $r$ under ordinary kriging (OK) and IRF universal kriging (UK) using the training data subset. 
``True $\kappa$'' represents the order from which the data were simulated. The $\hat{r}$ was determined using weighted least squares. The root-mean squared error (RMSE) was calculated by comparing predictions to the ``unobserved'' testing subset values.}
\label{tab:krg-rhat}
\begin{tabular}{c c c c c}
	\hline\noalign{\smallskip}
    	True $\kappa$ & Method & $\hat{\kappa}$ & $\hat{r}$ & RMSE \\
    	\noalign{\smallskip}\hline\noalign{\smallskip}
    	2 & UK & 2 & 0.717 & 0.07 \\
    	2 & OK & 1 & 0.888 & 10.71 \\
    	3 & UK & 3 & 0.707 & 0.07 \\
    	3 & OK & 1 & 0.992 & 47.19 \\
	\noalign{\smallskip}\hline
\end{tabular}
\end{table}

With the estimates of $\kappa$ and $r$, predictions were made for the 10\% ``unobserved'' locations corresponding to our testing dataset. The actual values in the testing dataset were compared against the predictions by calculating the root mean square error (RMSE). The RMSE between the universal kriging values and the actual values was only $0.07$ for both IRF2 and IRF3, respectively (Table \ref{tab:krg-rhat}). However, the ordinary kriging RMSE was approximately 11 and 47, respectively, showing a marked decrease in predictive capability when $\hat{\kappa}$ is incorrect. Naturally, the farther an IRF$\kappa$ process is from IRF1, the ability of ordinary kriging to make accurate predictions would be lower. \\


\section{Declarations}

The authors do not have any conflicts of interests. There was no funding for this research.

All code is available upon request.

For the author contributions, Drs. Shields and Huang led the development of the graphical estimation procedure. Drs. Bussberg and Huang led the development of the universal kriging procedure and associated simulation studies. Dr. Bussberg led the manuscript development.

\bibliography{irfkriging-bibliography}

\end{document}